\documentclass[preprint,12pt]{elsarticle}
\usepackage{graphicx}
\usepackage{bm}
\usepackage{amssymb}

\def\br{\begin{eqnarray}}
\def\er{\end{eqnarray}}
\def\be{\begin{equation}}
\def\ee{\end{equation}}

\begin{document}

\begin{frontmatter}
\title{Chiral symmetry breaking in QCD-like gauge theories with a confining propagator and dynamical gauge boson mass generation}
\author[ad]{A. Doff}
\ead{agomes@utfpr.edu.br}
\author[aan]{F. A. Machado}
\author[aan]{A. A. Natale}
\ead{natale@ift.unesp.br}
\address[ad]{Universidade Tecnol\'ogica Federal do Paran\'a - UTFPR,
Via do Conhecimento Km 01, 85503-390, Pato Branco - PR, Brazil }

\address[aan]{Instituto de F\'{\i}sica Te\'orica, UNESP - Universidade Estadual Paulista, Rua Dr. Bento T. Ferraz, 271, Bloco II, 01140-070, S\~ao Paulo - SP, Brazil}

\begin{abstract}
We study chiral symmetry breaking in QCD-like gauge theories introducing a confining {\it effective propagator}, as proposed recently
by Cornwall, and considering the effect of dynamical gauge boson mass generation. The effective confining propagator
has the form $1/(k^2+m^2)^2$ and we study the bifurcation equation finding limits on the parameter $m$ below which a satisfactory fermion mass
solution is generated. Considering the evidences that the coupling constant and the gauge boson propagator are damped in the infrared, due to the presence
of dynamically massive gauge bosons, the major part of the chiral breaking is mostly due to the confining propagator. We study the
asymptotic behavior of the gap equation containing confinement and massive gauge boson exchange, and find that the symmetry breaking can
be approximated at some extent by an effective four-fermion interaction generated by the confining propagator. We compute
some QCD chiral parameters as a function of $m$, finding values compatible with the experimental data. Within this 
approach we expect that lattice simulations should not see large differences between the confinement and chiral symmetry breaking 
scales independent of the fermionic representation and we find a simple approximate relation between the fermion condensate and dynamical
mass for a given representation as a function of the parameters appearing in the effective confining propagator. 
\end{abstract}
\end{frontmatter}

\section{Introduction}

QCD has two main properties: the chiral symmetry breaking (CSB) and confinement of quarks and gluons.
Both phenomena are related to the non-perturbative infrared (IR) dynamics. The dynamical generation
of quarks masses leading to CSB can be observed through the study of QCD Green functions, what has been extensively performed
in the last years by means of Schwinger-Dyson equations (SDE) \cite{robwil}. Green functions can also be studied through
simulations of gauge theories on the lattice, and, in particular, a great improvement in the understanding of dynamical
gauge boson mass generation in these theories was obtained recently as pointed out in Ref.\cite{cucchi}. In what concerns
gauge boson mass generation there is a nice agreement between the lattice computations and the solutions of Schwinger-Dyson 
equations \cite{aguilara,aguilarb}. The consistency between these two different approaches, strengthened by the phenomenological 
consequences \cite{natale}, reinforces the robust picture of gauge boson mass generation formulated by Cornwall
many years ago \cite{corna,cornb,cornc,cornd,corne,cornf}. In this scenario the dynamical gluon mass induce vortices in the theory 
and these may be responsible for confinement. 

The study of CSB through lattice simulations and the SDE approach have not come to the same firm ground
as in the case of gauge boson mass generation. First, early SDE studies of CSB have shown a minute
or absence of quark mass generation in the presence of dynamical gluon mass ($m_g(k^2)$) \cite{haeria,haerib,haeric}. Secondly,
in lattice simulations it seemed that CSB and confinement were triggered by the same mechanism (at least
for SU(2) theories \cite{bow1,holl1}). A recent result about CSB in SU(3) gauge theory indicates that we still need more simulations to confirm 
whether or not
confinement and CSB are intimately connected, but there is a clear sign of a connection between CSB and the string tension \cite{bowman}.
From the point of view of gap equations we may say that something is missing in the solution of the fermionic SDE,
and the most plausible possibility is confinement at least for fermions in the fundamental representation, 
although it may not be essential for CSB when fermions are
in higher dimensional representations (e.g. ``quarks" in the adjoint representation) \cite{cornwall2}. Two
recent papers tackle this problem in different ways: one introduces a confining effective  propagator into the gap 
equation \cite{cornwall3} and the other introduces a modified quark-gluon vertex containing information about the
ghost sector as well as makes use of lattice propagators \cite{aguilar2}. Both studies are able to explain CSB for
fundamental and adjoint quarks. One of them indicate that confinement may be an essential ingredient in
the gap equation \cite{cornwall3}. The other contains a set of effects which include the use of a gluon propagator obtained
from the lattice simulations and a complete vertex function that is enhanced in the infrared \cite{aguilar2}.

If CSB is a phenomena linked to confinement, as the lattice results seem to indicate, we may ask how the
confinement information is embodied into the lattice gluon propagator used in the fermionic gap equation in
order to generate quark masses. Anyhow, some gluon confinement effect ought to be present in the lattice propagator.
However it does not show the linear confining potential felt by quarks that has also been observed in the lattice
simulations \cite{greensite}. The linear potential that we are referring to is the successful phenomenological quark model potential
given by 
\be
V_F (r) = K_F r - \frac{4}{3} \frac{\alpha_s}{r} \,\, ,
\label{pot1}
\ee
where the confining first term is linear with the distance and proportional to the string tension $K_F$. The second term, that is 
of order $\alpha_s$, the strong coupling constant, describes the one gluon exchange contribution. The fact that the lattice
gluon propagator does not reproduce the confining part of this potential has been observed recently in Ref.\cite{vento}.
The potential between static quark charges is related to the Fourier transform of the time-time
component of the full gluon propagator in the following way
\be
V ({\bf{r}}) = - \frac{2C_2}{\pi} \int d^3 {\bf{q}} \alpha_s ({\bf{q}}^2) \Delta_{00}({\bf{q}}) \exp^{\imath {\bf{q.r}}} \,\, ,
\ee
where $C_2$ is the Casimir eigenvalue of the fundamental representation, the bold terms, ${\bf{q}}$ and ${\bf{r}}$, are 3-vectors.
$\Delta_{00}({\bf{q}})$ is the zero-zero component of the gluon propagator in the momentum configuration. As noticed in
Ref.\cite{vento} the linear confining term of the potential ($K_F r$) cannot be obtained from the gluon propagator 
determined in the lattice or from the gluonic SDE.  
Only the Fourier transform of a $1/q^4$ type of the product coupling$\otimes$propagator will generate a linear term in the quark potential,
although no fundamental field presents such behavior.

The proposal of Ref.\cite{cornwall3} is attractive because it includes the necessary
confining force into the gap equation, maintaining covariance and solves the problem discussed in the previous
paragraph. It is a simple procedure based on critical properties of confinement and, as
we will discuss, allows for a phenomenologically consistent discussion of CSB in gauge theories, providing a model
that may be tested through lattice calculations. The main idea is that entropic effects, that can be associated to an area law
for the Wilson quark loop, can be taken into account by a confining effective propagator of the form $1/(p^2+m^2)^2$ with a finite
parameter $m \propto K_F/M(0)$, where $M(0)$ is the zero-momentum value of the running quark
mass and $K_F$ is the string tension. Using this confining effective propagator, which is to be understood 
as the result of a collective effect not related to the propagator of a fundamental gluon field, and responsible for 
quark confinement. Given this proposal within its own reasons, we aim to investigate its consequences for CSB phenomenology
in the QCD case. The new gap equation is able to reproduce all QCD chiral symmetry breaking parameters.

We may say that the papers of Ref. \cite{aguilar2} and \cite{cornwall3} have complementary ideas and may even indicate
a possible mixed picture of the CSB mechanism. In Ref. \cite{aguilar2} it was noticed that even a quite complete treatment of the
vertex function and gauge dependence of the quark SDE does not allow for CSB in agreement with experiment. A gluon propagator
determined in the lattice, that is a little bit less damped in the intermediate infrared region when compared to the one obtained with
the gluon SDE, is necessary, as well as some strong ghost effects into the quark-gluon vertex. However, it is not clear that introducing 
some complicated ghost physics into the one-gluon gap equation would lead to confining effects typical of center vortices. It is also
questionable that the gluon propagators found on the lattice will be able to evidence confinement in, for example, a
Wilson loop. If so much confinement is present in the one-gluon-exchange approach it would not be simple to explain how the pion ends
up massless. Finally, it is not through open quark lines, as all gap equations use, that we shall observe confining effects, but through
closed loops. Hence there is certain appeal on the \textit{explicit} introduction of confining objects, such as center vortices. This is exactly
what Cornwall's proposal does \cite{cornwall3}, introducing an effective confining propagator whose effect
is to add an extra strength to the massive gluon exchange in the infrared, with the advantage of having a simple formula to model
the IR QCD behavior. We also recall that it has been argued that the one-gluon massive exchange and the confining gluon propagators act 
differently when quarks are in the fundamental and adjoint representations \cite{cornwall2}, therefore we have a precise way to investigate
the possible differences in the CSB and confinement transitions for fermions in different representations. 
We will argue that CSB is a direct consequence of confinement in this specific model.

The article is organized as follows. In Sect. II, we describe briefly the model of Ref.\cite{cornwall3}, i.e. how confinement is
introduced in the gap equation. In Sect. III, we study the bifurcation equation of the full gap equation. It is verified that
we have maximum $m$ values in order to generate reasonable values for the dynamical quark mass. In Sect. IV, we
present an analysis of the asymptotic behavior of the gap equation. The idea was to investigate how confinement could
affect the asymptotic behavior of the fermionic self-energy. It must be said that
the CSB mechanism is not only important for QCD, but for any non-Abelian gauge theory. In particular, it has consequences
for the Standard Model dynamical gauge symmetry breaking, or the so called Technicolor models. 
We find that if the confinement effective propagator were restricted to a limited region of momenta the self-energy
is well approximated by an effective four-fermion interaction. In Sect. V we compute quantities like the pion decay constant
and the quark condensate verifying that the model predicts reasonable values for the QCD chiral parameters. In Sect. VI
we determine an approximate relation between the fermion condensate and the dynamical
mass for a given representation as a function of the parameters appearing in the confining effective propagator.
These quantities can be studied through lattice simulations and may provide a test for this CSB mechanism.
In Sect. VII, we draw our conclusions.
 
\section{Introducing confinement into the gap equation}

The possible relation between confinement and chiral symmetry breaking for quarks in the fundamental representation
is an issue discussed several times in the literature in many different contexts as pointed out in Ref.\cite{cornwall2}. With the
evidence that gluons acquire a dynamically generated mass \cite{cucchi,aguilara,aguilarb}, i.e. a momentum dependent mass $m_g(k^2)$, it also becomes clear that the standard quark gap equation with dynamically massive gluons does not have enough strength to
generate quark masses. Therefore, if CSB and confinement are triggered by the same mechanism, the main quest in this subject is how to introduce confinement into the quark SDE.

The proposal of Ref.\cite{cornwall3} is that confinement should be introduced into the gap equation through the 
following {\it effective propagator}, {\it meaning that it is not at all related to the propagation of a standard quantum field}:
\be
D_{eff}^{\mu \nu}(k) \equiv \delta^{\mu \nu} D_{eff} (k); \,\,\,\,\,  D_{eff} (k)=\frac{8\pi K_F}{(k^2+m^2)^2}   \, ,
\label{eq01}
\ee
In the $m\rightarrow 0$ limit we obtain the standard effective propagator $8\pi K_F \delta^{\mu \nu}/k^4$, that yields approximately
an area law for the Wilson loop. We must necessarily have a finite $m\neq 0$ value due to entropic reasons as demonstrated in Ref.\cite{cornwall3}, and it is related to the dynamical quark mass, as
required by gauge invariance. Moreover, the Abelian gauge invariance of this effective
propagator must appear in the quark action obtained by integrating over quark world lines that will
imply a area-law action \cite{cornwall3}.  With the inverse fermionic propagator written as
\[
S^{-1}(p)= {\not\! p} A(p^2) + B(p^2) \,\, ,
\] 
using the approximation $A(p^2)=1$, which implies that $B(p^2)$ turns out to be identical to the dynamical mass $M(p^2)$, we
can see that the effective propagator of Eq.(\ref{eq01}) leads to an Abelian gluon gap equation equal to
\be
M_c(p^2)=\frac{1}{(2\pi)^4}\int \, d^4k \, D_{eff} (p-k) \frac{4M(k^2)}{k^2+M^2(k^2)} \,\, ,
\label{eq02}
\ee  
where $M_c(p^2)$ is the running dynamical quark mass generated by the confining propagator, so we end up with
\be
M_c(p^2)=\frac{1}{(2\pi)^4}\int \, d^4k \, \frac{8\pi K_F}{[(p-k)^2+m^2]^2} \frac{4M(k^2)}{k^2+M^2(k^2)} \,\, .
\label{eq0201}
\ee  
Note that there is an interplay between the parameters $K_F$, $m$ and $m_g$ as
extensively discussed in Ref.\cite{cornwall3}, and we will comment about their values, but it must be said that they should
be related among themselves because in the end all of them will be proportional to the QCD mass scale $\Lambda_{QCD}$.

\begin{center}
\begin{figure}[ht]
\hspace*{3cm}\includegraphics[scale=0.6]{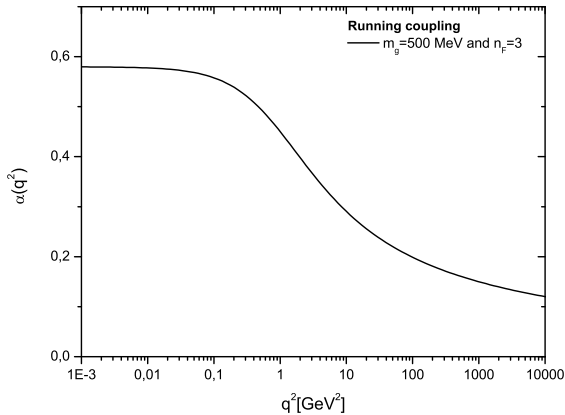}
\vspace{-0.1cm}
\caption{The running coupling constant for QCD with dynamically massive gluons.}
\label{fig01}
\end{figure}
\end{center}

Eq.(\ref{eq0201}) is not the whole story, since it was also recognized that massive one-dressed-gluon exchange may  
induce a quite small CSB for fermions in the fundamental representation as well as a larger symmetry breaking for fermions in
higher dimensional representations \cite{cornwall2,aguilar2}, which, in the Landau gauge, is given by
\be
M_{1g}(p^2) = \frac{C_2 }{(2\pi)^4}\int \, d^4k \,  \frac{{\bar{g}}^2(p-k)3M(k^2)}{[(p-k)^2+m_g^2(p-k)][k^2+M^2(k^2)]} \,\, ,
\label{eq03}
\ee
where $C_2$ is the quark Casimir eigenvalue and ${\bar{g}}^2$ is the effective charge
\be
{\bar{g}}^2(k^2)= \frac{1}{b \ln[(k^2+4m_g^2)/\Lambda_{QCD}^2]} \, ,
\label{eq04}
\ee
where $b=(11N-2n_f)/48\pi^2$ for the $SU(N)$ group with $n_f$ flavors. For quarks in the
fundamental representation $C_2 = 4/3$ and with the phenomenologically preferred value $m_g \approx 2\Lambda_{QCD} \approx 500-600$ MeV \cite{natale},
we see that this charge's value at the infrared fixed point \cite{natale2a,natale2b} ($\alpha_s (0)\equiv {\bar{g}}^2(0)/4\pi$) is of order $0.5$, as
shown in Fig.(\ref{fig01}), while
it should be at least a factor $2$ larger to trigger CSB.
For ``quarks" in the adjoint representation the Casimir eigenvalue
is approximately a factor two larger and compensates the small effective coupling \cite{cornwall2,aguilar2}.
The ultimate gap equation may be quite
sophisticated but it can be well modeled by the sum of the ``confining" plus massive one-gluon exchange \cite{cornwall3},
which, in the Abelian gluon approximation, is given by 
\br
M(p^2)&=&\frac{1}{(2\pi)^4}\int \, d^4k \, D_{eff} (p-k) \frac{4M(k^2)}{k^2+M^2(k^2)}   \nonumber \\
&&+\frac{C_2 }{(2\pi)^4}\int \, d^4k \,  \frac{{\bar{g}}^2(p-k)3M(k^2)}{[(p-k)^2+m_g^2(p-k)][k^2+M^2(k^2)]} \, ,
\label{eq0511}
\er
where $M(p^2)=M_c(p^2)+M_{1g} (p^2)$ is the dynamical quark mass generated by the confining and one-dressed-gluon propagators. This last equation is the basic one that we shall explore in this work. We could say that this equation resembles, in a different context, what we have
in the phenomenological quark model potential described in Eq.(\ref{pot1}), i.e. a part that is responsible by confinement
(generating an approximately linear term proportional to the string tension),
 and the second term, that is of order $\alpha_s$, describing the
one-gluon exchange contribution. Thereof, confining part of Eq.(\ref{eq0511}) is a reasonable phenomenological way to study
CSB taking into account the effective area law.

The solution of the confining gap equation (Eq.(\ref{eq0201})) was discussed by Cornwall \cite{cornwall3}, and it was observed
that the confining propagator generates CSB represented by a dynamical quark mass that is of order $M\approx m\approx \sqrt{K_F/\pi}\approx \Lambda_{QCD}$.
This dynamical quark mass has a fast large momentum falloff ($\propto 1/p^4$), which results from a gap equation basically dominated
by the infrared region of the propagator. On the other hand the $1$-gluon gap equation was extensively studied in the literature and it
is known that it generates CSB only above a certain critical coupling equal to $\alpha_c (0)\equiv (g^2_c/4\pi) \geq (\pi/3C_2)$, which,
as discussed in the previous paragraph, is not achieved by the effective coupling of Eq.(\ref{eq04}). Notice that the suggestion
of a hypothetical Casimir scaling law for CSB with fermions in the representation $R$ \cite{susa,susb}
\be
\alpha_s C_2(R) \approx {\cal{O}}(1)  \,\, ,
\label{cas1}
\ee
appears as a consequence of the SDE equation without the existence of a dynamical gauge boson mass.

It is easy to verify that Eq.(\ref{eq0201}) has nontrivial solutions. Its critical behavior can be inferred from the derivative of Eq.(\ref{eq03}) with respect to $m_g^2$, where we can replace $3{\bar{g}}^2 C_2$ by $8\pi K_F$ (in the constant coupling limit), linearize with the substitution of $M^2(k^2)$ in the denominator by $M^2$, and evaluate at $m_g=m$:
\be
\left. \frac{d M_{1g}(p^2)}{dm_g^2}\right|_{m_g = m} \propto M_c(p^2) \, \, .
\label{eq06}
\ee
Eq.(\ref{eq03}) has already been solved with different approximations \cite{haeria,haerib,cornwall3}, and has solutions
only for large ${\bar{g}}^2 C_2$ values as noticed in Ref.\cite{cornwall3}. The condition for Eq.(\ref{eq06}) to have a nontrivial
solution (and consequently Eq.(\ref{eq0201})) is the same one that Eq.(\ref{eq03}) has to obey. Considering the damped
confining propagator that leads to an effective coupling $8\pi (K_F/m^2)$ (that is now in the place of $3{\bar{g}}^2 C_2$), 
and the fact that $K_F \approx (3-4)m^2$ \cite{cornwall3}, 
we verify, just by transposing the results of Ref.\cite{haeria,haerib,cornwall3,atkinsona,atkinsonb} (see, particularly, the simple
analysis shown in section 3 of Ref.\cite{haerib}), that chiral symmetry will
always be broken with the confining propagator. Therefore, the above procedure leads to the solutions of Eq.(\ref{eq0201}) that can be represented by 
a linear combination of the following expressions:
\be
M_c^{\pm}(p^2)\approx M \theta (m^2-p^2) + M \left( \frac{2m^2}{p^2+m^2}\right)^{\gamma_\pm +1} \theta (p^2-m^2) \,\, ,
\label{eq07}
\ee
obtained from the derivative of the solutions of Eq.(\ref{eq03}), where $\theta$ is the step function and 
\[ 
\gamma_\pm = \frac{1}{2}\left[1\mp (1-4\lambda )^{1/2}\right] \,\,\,\,\, , \,\,\,\,\, \lambda = K_F/2\pi \,\, . 
\]
The value of $M$ is determined by the
boundary condition at $p^2=0$
\be
M\equiv M_c(p^2=0)=\frac{2K_f}{\pi} \int_0^\infty \, dk^2 \, \frac{k^2M(k^2)}{(k^2+m^2)^2 (k^2+M^2)} \,\, .
\label{eq08}
\ee

It seems that the confining effective propagator for the $K_F$ and $m$ values discussed above is able to describe the
chiral symmetry breaking. However the full CSB problem, as discussed by Cornwall \cite{cornwall3}, includes the effect of the
dressed $1$-gluon exchange, and the critical behavior for the onset of dynamical quark masses in this case will be
discussed in the next section. It is possible that only the sum of these effects may explain the lattice results for CSB \cite{bowman}.

\section{Critical behavior of the complete gap equation}

We can discover some aspects about Eq.(\ref{eq0511}) critical behavior examining its
bifurcation equation. To verify at what point the nontrivial solution of Eq.(\ref{eq0511}) bifurcates away from trivial solution, 
it is sufficient to consider the linearized version of that equation \cite{atkinsona,atkinsonb}. We will deviate from the standard
bifurcation theory proceeding as in Ref.\cite{atkinson0}, and instead of substituting $k^2+M^2(k^2)$ by $k^2$ in the denominators
of Eq.(\ref{eq0511}), we will replace this term by $k^2+\delta M^2(0)$ and define the IR value of the dynamical quark mass ($M$) by the
normalization condition
\[
\delta M(0)= M \, \, ;
\]
following these steps we arrive at our bifurcation equation
\br
\delta M(p^2)&=&\frac{1}{(2\pi)^4}\int \, d^4k \,  \frac{8\pi K_F}{[(p-k)^2+m^2]^2}  \frac{4\delta M(k^2)}{k^2+M^2} \nonumber \\ 
&&+\frac{C_2 }{(2\pi)^4}\int \, d^4k \,  \frac{{\bar{g}}^2(p-k)3 \delta M(k^2)}{[(p-k)^2+m_g^2][k^2+M^2]} \, .
\label{eq051}
\er

Our main intention in this section is to verify the gross critical behavior of the gap equation with these infrared
finite propagators, this is the reason to select a bare massive gluon in Eq.(\ref{eq051}). This equation is
a standard Fredholm equation with a positive kernel, and, requiring $\delta M(p^2)$ to belong to $L^2$, the 
spectrum is discrete with a smallest value for the ``effective coupling" $K_F/m^2$ and the $1$-gluon exchange
coupling ${\bar{g}}^2/4\pi$ such that we have the trivial solution $\delta M(p^2)\equiv 0$ for values of these couplings smaller than a certain critical value, and the nontrivial one if their values are larger than this
same critical value.

We can still make some simplifier approximations before estimating the critical behavior of Eq.(\ref{eq051}),
making the following substitutions
\be
\alpha \left[ (p-k)^2/\Lambda_{QCD}^2 \right]\equiv  \frac{{\bar{g}}^2[(p-k)^2]}{4\pi}\rightarrow
\theta (p^2-k^2) \alpha (p^2) + \theta (k^2-p^2) \alpha (k^2) \,\, ,
\ee
and
\be
\frac{1}{(p-k)^2+m_g^2} \rightarrow \frac{1}{p^2+m_g^2}\theta (p^2-k^2)+\frac{1}{k^2+m_g^2}\theta (k^2-p^2) \,\, ,
\ee
which is known as the angle approximation, and introduces an error of about $10\%$ in the calculation \cite{rob}.
If we also define the variables $x=p^2/M^2$, $y=k^2/M^2$, $\kappa = m^2/M^2$, $\epsilon = m_g^2/M^2$,
$\rho = \Lambda_{QCD}^2/M^2$ and $f(p^2)= \delta M (p^2)/M$, we obtain 
\be
f(x)=\frac{1}{\pi}\int_0^{\Lambda^2/M^2} \, dy \, K(x,y) f(y) \,\, ,
\label{eq09}
\ee
where we introduced an ultraviolet cutoff ($\Lambda$) and the kernel $K(x,y)$ is equal to
\br
K(x,y)&=&  \frac{y}{(y+1)} \left[ \left( \frac{2K_f}{M^2}\frac{1}{(y+\kappa )^2} +\frac{3C_2}{16\pi} \right. \right.
\left. \frac{{\bar{g}}^2(y)}{(y+\epsilon)} \right) \theta (y-x) \nonumber \\
&+& \left( \frac{2K_f}{M^2}\frac{1}{(x+\kappa )^2} +\frac{3C_2}{16\pi} \right.
\left. \left. \frac{{\bar{g}}^2(x)}{(x+\epsilon)} \right) \theta (x-y)\right] \,\, .
\label{eq10}
\er

The kernel $K$ is square integrable
\br
\left\| K \right\|^2 &=& \int_0^{\Lambda^2/M^2} \, dx \, \int_0^x \, dy \,
 \frac{y^2}{(y+1)^2}\left( \frac{2K_f}{M^2}\frac{1}{(x+\kappa )^2} +\frac{3C_2}{16\pi} \right.
\left. \frac{{\bar{g}}^2(x)}{(x+\epsilon)} \right)^2 \nonumber \\
&+& \int_0^{\Lambda^2/M^2} \, dx \, \int_x^{\Lambda^2/M^2} \, dy \,
\frac{y^2}{(y+1)^2}\left( \frac{2K_f}{M^2}\frac{1}{(y+\kappa )^2} +\frac{3C_2}{16\pi} \right.
\left. \frac{{\bar{g}}^2(y)}{(y+\epsilon)} \right)^2 \,\, ,
\label{eq11}
\er
therefore Eq.(\ref{eq09}) has a nontrivial $L^2$ solution for its coupling on a point set.
The smallest eigenvalue (which, as we shall see, is related to the value of $K_F/M^2$) for which
Eq.(\ref{eq09}) has a nontrivial square integrable solution, is the first bifurcation of the 
nonlinear equation, and satisfies
\be
\frac{1}{\pi} \left\| K \right\| = 1 \,\, .
\label{eq12}
\ee
Note that the kernel $K$ contains the sum of two contributions, that we may denote $K_c$ and $K_{1g}$, corresponding to the confining
and dressed $1$-gluon exchange propagator.
From the triangle inequality we know that
\be
\left\| K_c +K_{1g} \right\| \leq \left\| K_c \right\| + \left\| K_{1g} \right\| \,\, ,
\label{eq13}
\ee
from which we can recover the early results for the $1$-gluon exchange, i.e. if the gluon mass is not
introduced into the gap equation and the confining propagator is neglected, the limit of Eq.(\ref{eq12}) is obtained only due to the $K_{1g}$
contribution with a critical coupling constant $\alpha_c \approx {\cal{O}}(1)$, but the $K_c$ contribution is necessary when considering dynamically 
massive gluons and quarks in the fundamental representation. 

\begin{center}
{\bf \begin{figure}[ht]
\hspace*{3.8cm}\includegraphics[scale=0.8]{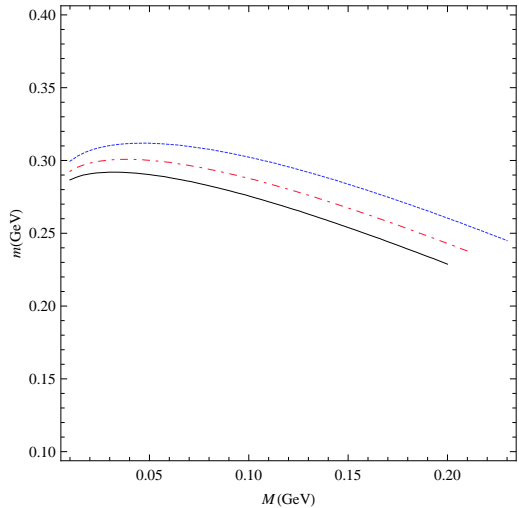}
\vspace{-0.1cm}
\caption{The criticality condition given by Eq.(\ref{eq12}) for the kernel of Eq.(\ref{eq10}) is plotted in the case of $n_f =3$,
$\Lambda_{QCD} =300$MeV and $K_F = 0.18$GeV$^2$. The dashed (blue) curve was obtained with $m_g=600$MeV, the dot-dashed (red) curve
with $m_g=650$MeV and the solid (black) curve with $m_g=700$MeV. }
\label{fig1}
\end{figure}}
\end{center}

\begin{center}
\begin{figure}[ht]
\hspace*{3cm}\includegraphics[scale=0.55]{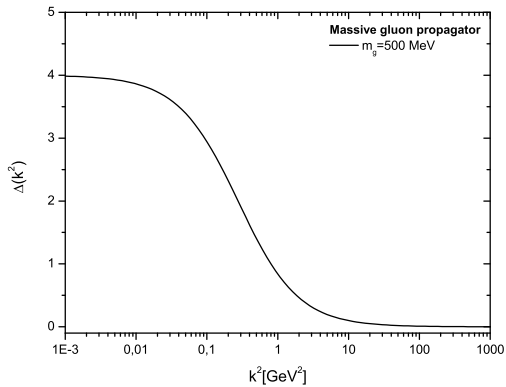}
\vspace{-0.1cm}
\caption{The massive gluon propagator with $n_f =3$ and
$\Lambda_{QCD} =300$ MeV. }
\label{fig02}
\end{figure}
\end{center}

\begin{center}
\begin{figure}[t]
\hspace*{3cm}\includegraphics[scale=0.6]{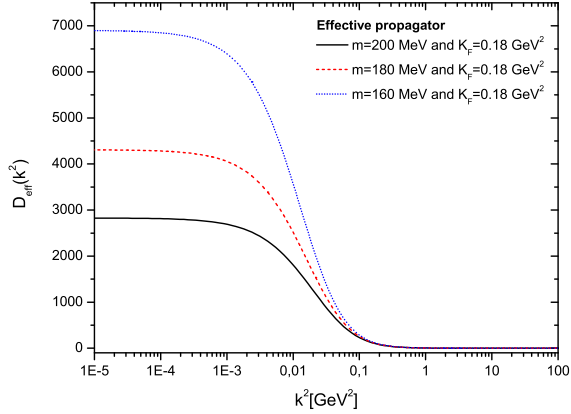}
\vspace{-0.1cm}
\caption{The confining effective propagator of Eq.(\ref{eq01}) with $K_F = 0.18$ GeV$^2$ and different
$m$ values. Note the ${\cal{O}}(10^3)$ strength difference when compared to Fig.(\ref{fig02}).}
\label{fig03}
\end{figure}
\end{center}

The bifurcation condition described by Eq.(\ref{eq12}) is depicted in Fig.(\ref{fig1}) in the case of $n_f =3$,
$\Lambda_{QCD} =300$MeV and $K_F = 0.18$GeV$^2$, where it is shown a dashed (blue) curve obtained with $m_g=600$MeV, a dot-dashed (red) curve
with $m_g=650$MeV and a solid (black) curve with $m_g=700$MeV. Each point of these curves indicate the bifurcation point for a given
$m$ value generating a dynamical quark mass $M$. It should be noticed that there is a maximum $m$ value above which there is no CSB,
and this maximum value does not vary much as we change the values of the dynamical gluon mass. It is also interesting 
to verify that CSB also receives contributions from the massive gluon term, and this is the reason for the differences between the curves,
though the massive gluon (or $1$-gluon exchange) generates a minute mass, as already observed many years ago in Ref.\cite{haeria,haerib}. As
the dynamical gluon mass is decreased we can observe a small increase in the maximum value of the $m$ parameter, due to the 
$1$-gluon exchange increasing contribution to the gap equation.
When the bifurcation condition is computed it is possible to verify a larger instability in the numerical procedure as we go to larger values 
of $m$ and $m_g$ (for large $M$ values), which is due to the fact that we have two contributions to the
chiral breaking, but with one of the kernels contributing much more to the breaking than the other. This fact is obvious if we
compare the different propagators that we are dealing with. These propagators are shown in Fig.(\ref{fig02}) and Fig(\ref{fig03}) for
typical mass parameters that we are discussing in this work. We can see a huge difference between 
the two propagators (${\cal{O}}(10^3)$). Of course, the confining effective propagator 
should not be regarded as the actual gluon propagator, but must be seen as a collective effect that produces confinement,
with an area law for the quark action and appropriate entropic properties \cite{cornwall3}. The difference between these
contributions are responsible for the delicate convergence of the bifurcation condition.

It must be also noticed that the confining effective propagator has most of its effect concentrated in a momentum region 
below ${\cal{O}}(100)$ MeV. Actually, as we shall discuss ahead, most of the chiral symmetry breaking will occur due
to the low momentum region of the gap equation, what is consistent with lattice observations that the relevant
momentum component of gluons for CSB is exactly this region \cite{sugaa,sugab,sugac}. This fact may be contrasted with
the results of Ref.\cite{aguilar2}, where a larger part of the CSB comes from an intermediate region of ${\cal{O}}(1)$ GeV.

\vspace{-2cm}

\section{The asymptotic behavior of the gap equation}

The asymptotic behavior of the complete gap equation [Eq.(\ref{eq0511})] can be obtained from the linearized Eq.(\ref{eq09}) with the kernel
given by Eq.(\ref{eq10}) in
a procedure identical to the one performed by Takeuchi \cite{takeuchi} and Kondo, Shuto and Yamawaki \cite{kondo}. In these references
the QCD CSB problem was solved considering a quark Schwinger-Dyson equation also with a two kernel contribution: One due to an effective four-fermion
interaction and another one due to a perturbative gluon exchange. 

The integral equation (\ref{eq09}) with the kernel described in Eq.(\ref{eq10}) can be transformed into a differential equation and
solved with appropriate boundary conditions. In order to still simplify the calculation we assume
\be
\kappa = \epsilon = \rho = 1 \, \, ,
\label{eq13a}
\ee
and neglect the running of the coupling constant in Eq.(\ref{eq10}). The fact that we consider all masses equal to $M$, as implied
by Eq.(\ref{eq13a}), shall not modify the
asymptotic behavior of the quark self-energy, and the assumption of a constant coupling will introduce sub-leading
or logarithmic corrections to the asymptotic solution. After these approximations we obtain the following differential equation 
\br
(x+1)^3[2a_1 (x+1)&+&a_2(x+1)^2]f^{\prime\prime}(x)+(x+1)^3[6a_1+2a_2(x+1)]f^\prime (x) \nonumber \\
&+& x[2a_1+a_2(x+1)]^2 f(x)=0 \,\, ,
\label{eq14}
\er
where
\be
a_1=\frac{2K_f}{\pi M^2} \,\,\,\,\,\, , \,\,\,\,\,\, a_2=\frac{3C_2 {\bar{g}}^2}{16 \pi^2} \,\,\, ,
\label{eq15}
\ee
with the infrared and ultraviolet boundary conditions given respectively by
\be
\left. \left. f(x)\right|_{x\rightarrow 0} = 1 \,\,\,\,\,\, , \,\,\,\,\,\, f^\prime(x)\right|_{x\rightarrow \Lambda^2/M^2} = 0 \,\, .
\label{eq16}
\ee

The asymptotic solution of the linear second-order differential equation, with a singularity at infinity, is obtained as a linear combination of two
independent solutions, $f(x)= b_1 f_+(x) + b_2 f_-(x)$, which can be obtained by applying the expansion method \cite{olver}. Eq.(\ref{eq14})
can be put in the form $f^{\prime\prime}(x)+p(x)f^\prime +q(x)f(x)=0$, where $p(x)$ and $q(x)$ can be expanded in convergent
series of the form $p(x)=(1/x) \sum_{s=0}^{\infty} p_s x^{-s}$ and $q(x)=(1/x^2) \sum_{s=0}^{\infty} q_s x^{-s}$ for large $x$. The
asymptotic solutions are of the form
\be
f_i (x) = \frac{1}{x^{\alpha_i}} \sum_{s=0}^{\infty} \frac{c^i_s}{x^s} \,\, ,
\label{eq17}
\ee
where $\alpha_i$ are the roots of
\be
\alpha_i(\alpha_i +1) - p_0 \alpha_i + q_0 =0 \,\, .
\label{eq18}
\ee
With the leading coefficients of the differential equation given by $p_0 = 2$ and $q_0 = a_2$ we obtain two roots
\be
\alpha_\pm = \frac{1\pm \sqrt{1-4a_2}}{2} = \frac{1\pm \omega}{2}\,\, .
\label{eq19}
\ee
Defining a new variable
\be 
\gamma_{\pm} = \frac{\omega \pm 1}{2}\,\,\, ,
\label{eq20}
\ee 
\noindent we can write the asymptotic solution as
\be 
f(x) = b_1 x^{-\gamma_{+}}\sum^{\!\!\infty}_{{\!\!s=0}}\frac{c^+_s}{x^s}   +  b_2 x^{\gamma_{-}}\sum^{\!\!\infty}_{{\!\!s=0}}\frac{c^-_s}{x^s}
\label{eq21}
\ee 
The substitution of Eq.(\ref{eq17}), with $\gamma_{\pm}$ given by Eq.(\ref{eq20}), into the differential equation (\ref{eq14}) allows the determination of the $c^i_s$ coefficients and the recursion formula satisfied by them.

Considering only the ``$+$" solution displayed in Eq.(\ref{eq21}), with the definitions
\br
&& c^{''}_{s} =  c^+_s(\gamma_{+} + s + 1)(\gamma_{+} + s ) \,\, ,  \nonumber  \\ 
&& c^{'}_{s} =  c^+_s(\gamma_{+} + s ) \,\, , \nonumber \\
&& c_{s}  = c^+_s  \,\, ,
\label{eq211}
\er
we obtain the following recursion formula
\br 
&& a_2c^{''}_{s+3} + (2a_1 + 5a_2)a_2c^{''}_{s+2} + (8a_1 + 10a_2)c^{''}_{s+1} + (12a_1 + 10a_2)c^{''}_{s}\nonumber \\ 
&&  + a^2_2c_{s+3} + 2a_2(a_2 + 2a_1)c_{s+2} + (4 (a^2_1 + a_1a_ 2) + a^2_2)c_{s+1}  - 2a_2c^{'}_{s+3} \nonumber \\
&&  - (8a_2 + 6a_1)c^{'}_{s+2} - (12a_ 2 + 18a_1)c^{'}_{s+1} - (8a_ 2 + 18a_1)c^{'}_{s}  = 0  \,\, ,
\label{eq212}
\er 
and verify, for example, that the first two coefficients of the ``$+$" series are related by
\br 
c^+_1 = -\frac{( 2a_2(a_2 + 2a_1) + (2a_1 + 5a_2)a_2(\gamma_{+} + 1) \gamma_{+} - \gamma_{+}(8a_2 + 6a_1))}{a^2_2  + a_2(\gamma_{+} + 1)^2 - a_2(\gamma_{+} + 1)}c^+_0 \,\, .
\label{eq213}
\er
For the ``$-$" solution we just have to exchange $c^+$ for $c^-$ and $\gamma_{+}$ for $-\gamma_{-}$ in Eqs.(\ref{eq211}) to (\ref{eq213}).
If we keep only the leading terms in $f_\pm (x)$ we can write $b_1 f_+ (x) \approx b_1 x^{-\gamma_{+}}(1+c_1^+ /x)\equiv f_{reg}^{asymp}$ and 
$b_2 f_- (x) \approx b_2 x^{\gamma_{-}}(1+c_1^- /x)\equiv f_{irreg}^{asymp}$. Applying the UV boundary condition we obtain that
\be
-\frac{b_1}{b_2}=x^\omega
\frac{\left(\frac{c_1^-}{x}(1-\gamma_-)-\gamma_-\right)}{\left(\frac{c_1^+}{x}(1+\gamma_+)+\gamma_+)\right)} \,\, .
\label{eq22}
\ee
The IR boundary conditions force the solutions to be equal to $1$ as $x\rightarrow 0$, and the amount each solution ($f_+$ or $f_-$) contributes 
to the self-energy can be measured by the ratio $R$ 
\be
R \propto \left. \frac{f_{reg}^{asymp}}{f_{irreg}^{asymp}}\right|_{x\rightarrow \Lambda^2/M^2} \,\, .
\label{eq23}
\ee

The ratio $R$ indicates which solution dominates the CSB. To gain some insight on the problem we can
recall that when the gluons are massive the coupling constant (see Eq.(\ref{eq04})) is not expected to be larger than $0.5$,
therefore $a_2$ is a small number, and we can see that in the limit $a_2\approx 0$, the coefficients $c_1^+$ and $c_1^-$ are
large and $\omega=\sqrt{1-4a_2}\approx 1$, leading us to $\gamma_- =\frac{\omega -1}{2}\approx 0$ and $\gamma_+ = \frac{\omega +1}{2}\approx 1$,
implying
\be
R \approx \left. \frac{c_1^-}{2c_1^+}x^\omega\right|_{x\rightarrow \Lambda^2/M^2}  \,\, .
\label{eq24}
\ee
In the ultraviolet limit we verify that the ratio $R$ behaves as $(\Lambda^2/M^2)^\omega$ and the $f_{reg}$ solution gives
the asymptotic behavior of the self-energy. This is, apart from logarithmic contributions, the known $1/p^2$ behavior found
by Politzer using the operator product expansion \cite{politzer}. Notice that the asymptotic behavior is fully described
by the $1$-gluon exchange, and this is not much different from what was found by Takeuchi \cite{takeuchi} in a problem where 
the CSB is dominated by a four-fermion interaction. The influence of the confining propagator enters only through the boundary 
conditions. As the mass solution that comes from the confining contribution has a fast falloff, almost behaving as an effective 
four-fermion interaction it is not surprising at all to see the similarity of our Eq.(\ref{eq24}) to Eq.(29) of Ref.\cite{takeuchi}.
Another interesting limit of Eq.(\ref{eq24}) is obtained when $\omega \rightarrow 0$, or $4a_2\equiv 3C_2 {\bar{g}}^2 /4\pi^2\approx 1$,
although this may happen only for almost conformal theories, and it is a case of interest for technicolor models. 
We mention this because ${\bar{g}}^2$ has an upper limit
when the gauge bosons acquire a dynamically generated mass, as shown in Eq.(\ref{eq04}) \cite{cornwall4}, and consequently we must
force the $\beta$ function coefficient $b$ in Eq.(\ref{eq04}) to be very small in order to have $\omega \rightarrow 0$, 
and it may be quite difficult to build realistic technicolor
theories with $4a_2 \approx 1$. This particular possibility is under study and will be presented elsewhere.
\begin{center}
\begin{figure}[ht]
\hspace*{3.5cm}\includegraphics[scale=0.6]{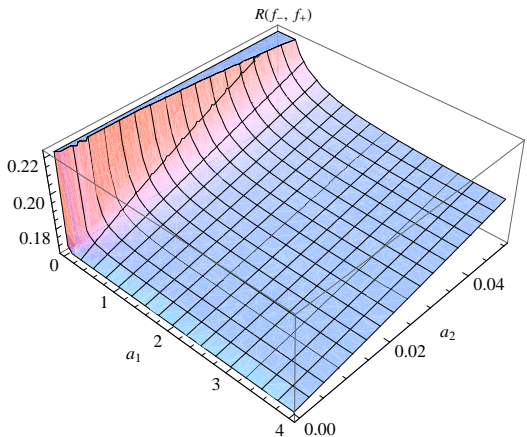}
\vspace{-0.1cm}
\caption{The ratio $R =f_{reg}^{asymp}/f_{irreg}^{asymp} $ when $\Lambda^2/M^2 \approx m^2/M^2 \approx 1$ as a function of some natural values for the parameters $a_1$ and $a_2$, which are proportional
to the effective couplings for the different contributions entering into the gap equation. In this particular case the figure indicates that
the irregular solution dominates over the regular one for a larger coupling value $(\propto a_1)$ of the confining propagator.}
\label{fig2}
\end{figure}
\end{center}

Equation (\ref{eq24}) would have a different (and interesting) behavior if the upper cutoff ($\Lambda$) were of 
order $m\approx M$. This would happen if the integration of the confining part of the gap equation, the one responsible
for the CSB, were limited to a region in momentum smaller than, or of the order of the scale $m$. 
Assuming an upper limit in the momentum integration,
we show in Fig.(\ref{fig2}) the ratio $R$ when $\Lambda \approx m$ as a function of $a_1$ and $a_2$. This ratio is smaller than $1$ indicating that
the asymptotic behavior of the irregular symmetry breaking solution would dominate over the regular one in this particular limit. 
In order to explore even more this possibility we can assume that the confining contribution could be reduced to an effective 
four-fermion interaction. Some of the reasons why we are concerned with the possibility of generating a four-fermion interaction are the following: 
First, if confinement is introduced into the gap equation we should
expect to reproduce some of the many phenomenological successful quark-models based on the Nambu-Jona-Lasinio type of interaction. Secondly,
lattice simulations show that the relevant gluonic energy scale of spontaneous CSB is due to the low-momentum component of the gluon field \cite{sugaa,sugab,sugac},
which may indicate the possibility of a natural upper cutoff in the momentum. The existence of a specific momentum that separates
the confinement and perturbative regions has also been discussed in a different context \cite{brodsky}.
Finally, the existence of a completely nonperturbative infrared fixed point, as happens when the theory develops a dynamical
gauge boson mass \cite{natale2a}, may induce effective four-fermion interactions as discussed many years ago in Ref.\cite{barda,bardb}.

It is known that 
as long as we have a ``massive" gluon propagator it would be possible to consider that this mass could be factorized from
the propagator generating an effective four-fermion interaction, but this is not true because the actual interaction strength is
measured by the product ``coupling$\otimes$propagator", and we know from Eq.(\ref{eq04}) that the $1$-gluon exchange has not
strength enough to generate such effective coupling. On the other hand the confining effective propagator, with the usual values for
the string tension, is strong enough to generate the following effective gap equation:
\br
M_{4f}(p^2)&=&\frac{2}{\pi^3} \frac{K_F}{m^4} \int \, d^4k \,  \frac{M_{4f}(k^2)}{k^2+M_{4f}^2(k^2)} \theta (m^2-k^2) \nonumber \\
&+&\frac{C_2 }{(2\pi)^4}\int \, d^4k \,  \frac{{\bar{g}}^2(p-k)3M_{4f}(k^2)}{[(p-k)^2+m_g^2(p-k)][k^2+M_{4f}^2(k^2)]} \,\, .
\label{eq25}
\er
Apart from the gluon mass effect appearing in the $1$-gluon contribution, the above equation has been extensively 
studied in Refs. \cite{takeuchi} and \cite{kondo}, and it does lead to a self-energy solution that decreases slowly with
the momentum, or the so called irregular solution for the self-energy \cite{kondo}. The solution of Eq.(\ref{eq25}) 
follows from Refs. \cite{takeuchi} and \cite{kondo} observing the
interplay between their $4$-fermion coupling constant $\lambda$ and our effective coupling $K_F /m^2$.

The critical behavior of Eq.(\ref{eq25}) can be compared to the one of the complete Eq.(\ref{eq09}) studied
in the previous section, but now with a $4$-fermion effective kernel given by 
\br
\left\| K_{4f} \right\|^2 &=& \int_0^\infty \, dx \, \int_0^x \, dy \,
 \frac{y^2}{(y+1)^2}\left(\frac{2K_f}{M_{4f}^2\kappa^2} \theta (\kappa -y)+ \frac{3C_2}{16\pi} \right.
\left. \frac{{\bar{g}}^2(x)}{(x+\epsilon)} \right)^2 \nonumber \\
&+& \int_0^\infty \, dx \, \int_x^\infty \, dy \, 
\frac{y^2}{(y+1)^2}\left(\frac{2K_f}{M_{4f}^2\kappa^2}\theta (\kappa -y) + \frac{3C_2}{16\pi} \right.
\left. \frac{{\bar{g}}^2(y)}{(y+\epsilon)} \right)^2 
\label{eq26}
\er

We separated the kernel of Eq.(\ref{eq26}) into two different kernels, one due to the effective four-fermion
interaction and another due to the exchange of a massive gluon. Using the triangle inequality of Eq.(\ref{eq13})
and the bifurcation condition given by Eq.(\ref{eq12}) we compute the critical condition and show in Fig.(\ref{fig3}) the dot-dashed (black) curve
of critical $m$ values for the generation of massive solutions of Eq.(\ref{eq25}). This curve was obtained for $m_g = 600$ MeV, $n_f =3$,
$\Lambda_{QCD} =300$ MeV and $K_F = 0.18$ GeV$^2$, and for comparison we also draw in Fig.(\ref{fig3}) the dashed (blue) critical curve of the
complete kernel given by Eq.(\ref{eq10}) (without the four-fermion approximation) 
computed with the same parameters. This shows that most of the symmetry breaking 
is driven by the confining effective
propagator and the four-fermion approximation is reasonable up to an order of $10\%$. Since we simplified the gap equation
and used the triangle inequality to obtain the curve of Fig.(\ref{fig3}), it is difficult to say how much of the difference
between these curves result from the numerical procedure. However it indicates that the confinement effect, as proposed in
Ref.\cite{cornwall3}, may indeed be the generator of an effective four-fermion interaction.

\begin{center}
{\bf \begin{figure}[ht]
\hspace*{3.8cm}\includegraphics[scale=0.8]{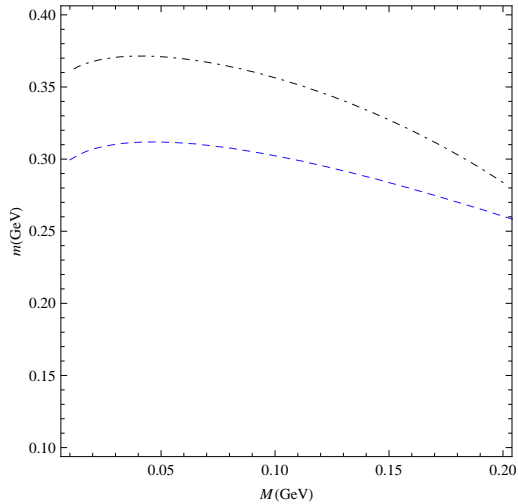}
\vspace{-0.1cm}
\caption{The bifurcation condition given by Eq.(\ref{eq12}) in association with the triangle inequality of Eq.(\ref{eq13}) is
used to compute the critical $m$ values for the massive solutions of Eq.(\ref{eq25}). The result for $m_g = 600$ MeV, $n_f =3$,
$\Lambda_{QCD} =300$ MeV and $K_F = 0.18$ GeV$^2$ is shown by the dot-dashed (black) curve. For comparison we also draw the dashed 
(blue) critical curve of the complete kernel given by Eq.(\ref{eq10}) computed with the same parameters.}
\label{fig3}
\end{figure}}
\end{center}

We also computed the bifurcation condition (Eq.(\ref{eq12})) with the part only due to the confining kernel in the four-fermion
approximation for different values of the string tension. The result is shown in Fig.(\ref{fig3l}) where the continuous
(blue) curve was obtained for $K_F = 0.18$ GeV$^2$ and the dot-dashed (black) and dashed (blue) were obtained respectively
for $K_F = 0.20$ and $K_F = 0.25$ GeV$^2$. Small changes in the string tension introduce larger effects than the ones
that can be observed when we vary the gauge boson mass in the bifurcation case of the complete gap equation. This confirms
again the dominance of the confining effective propagator.

There is no doubt that the confining gap equation can be reduced to an effective four-fermion approximation. However we
would like to claim that the confining part of the fermionic SDE should have an upper cutoff at some scale not too much
different from $m$. The reason for this is that the linear potential must break at some critical distance. For $n_f =2$ quarks in the fundamental representation, lattice QCD data shows that the string breaks at the following critical distance \cite{bali} 
\be
r_c \approx 1.25 \,\,\, fm \,\, ,
\label{critm}
\ee
which corresponds to a $m$ value compatible with the one necessary for the expected amount of CSB.

\begin{center}
{\bf \begin{figure}[ht]
\hspace*{3.8cm}\includegraphics[scale=0.8]{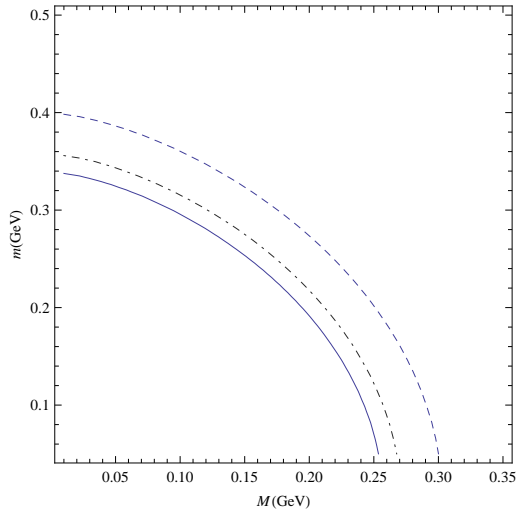}
\vspace{-0.1cm}
\caption{The  bifurcation condition (Eq.(\ref{eq12})) is plotted only with the contribution of the confining kernel in the four-fermion
approximation for different values of the string tension. The continuous
(blue) curve was obtained for $K_F = 0.18$ GeV$^2$ and the dot-dashed (black) and dashed (blue) were obtained respectively
for $K_F = 0.20$ and $K_F = 0.25$ GeV$^2$.}
\label{fig3l}
\end{figure}}
\end{center}
The results obtained up to now contain many approximations, as the assumption that all mass scales were the same. A realistic calculation 
taking into account the different mass scales that we have in the original equation will be performed numerically in the next section, however 
we may see that the effect of the confining propagator dominates over the $1$-gluon exchange and may even modify the asymptotic
behavior of the gap equation as seen in the four-fermion approximation, which retains the CSB information.

\section{Chiral parameters with confinement and massive gluons}

In the above sections we studied aspects of the bifurcation equation as well as the asymptotic behavior of the full gap equation, but
to know how the proposal of Ref.\cite{cornwall3} provides the right amount of chiral symmetry breaking it is necessary a
full computation of the complete gap equation without the approximations made above or in Ref.\cite{cornwall3}. We now solve
Eq.(\ref{eq0511}) without the angle approximation and taking into account the running coupling constant, with its infrared value
dictated by a dynamical gluon mass, and a massive gluon propagator with the running mass included exactly as determined in the
Ref.\cite{corna}. We choose the
same values of $n_f$, $\Lambda_{QCD}$, and $K_F$ that were used to obtain Fig.(\ref{fig1}), but select an infrared value
of the dynamical gluon mass equal to $m_g=500$ MeV, which is an average of many determinations of this parameter \cite{natale}. 
We compute numerically the full gap equation for the dynamical quark mass $M(p^2)$ with different $m$ values and 
show the results in Fig.(\ref{fig4}). 

\begin{center}
{\bf \begin{figure}[ht]
\hspace*{3cm}\includegraphics[scale=0.6]{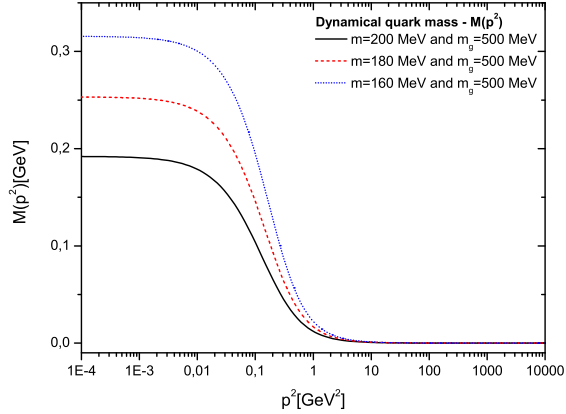}
\vspace{-0.1cm}
\caption{Dynamical quark mass obtained with the numerical calculation of the full gap equation (Eq.(\ref{eq0511})).}
\label{fig4}
\end{figure}}
\end{center}
A reasonable dynamical quark mass of ${\cal{O}}(250)$ MeV is obtained with a
$m$ value equal to $180$ MeV. This is a little smaller but totally consistent, within the different approximations, to the one
obtained in Ref. \cite{cornwall3}.

In order to show how reasonable the four-fermion approximation is to the complete gap equation we plot in Fig.(\ref{fig5})
the numerical calculation of Eq.(\ref{eq25}). This figure was obtained with $m=180$ MeV, $n_f =3$,
$\Lambda_{QCD} =300$ MeV, $K_F = 0.18$ GeV$^2$ and $m_g=500$ MeV. The infrared value of the dynamical quark mass is overestimated by an
${\cal{O}}(40\%)$ compared to the one of Fig.(\ref{fig4}), while it is also evident the slowly decreasing behavior
of the dynamical fermion mass with the momentum as it was advanced in the previous section. The fact that the four-fermion approximation overestimates
the dynamical quark mass was already observed in Fig.(\ref{fig3}). 
\begin{center}
{\bf \begin{figure}[ht]
\hspace*{3cm}\includegraphics[scale=0.6]{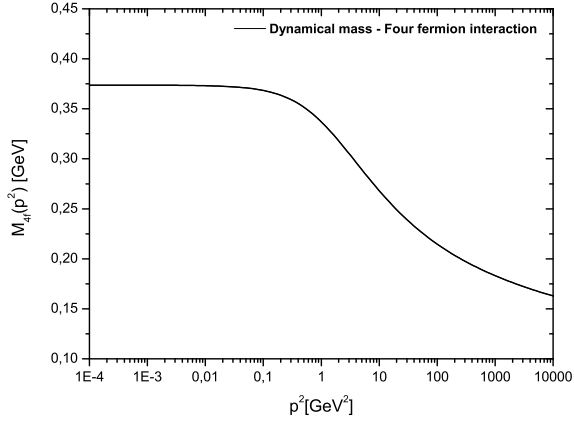}
\vspace{-0.1cm}
\caption{Dynamical quark mass obtained with the numerical calculation of the gap equation within the four-fermion
approximation (Eq.(\ref{eq25})).}
\label{fig5}
\end{figure}}
\end{center}

To confirm that the scenario of Ref. \cite{cornwall3} is fully consistent with the CSB phenomenology, we compute several
chiral parameters with the same values of $n_f$, $\Lambda_{QCD}$, $K_F$ and $m_g$ discussed above and $m=180$ MeV, which leads 
to the usually expected dynamical quark mass of $250$ MeV. The chiral parameters, computed in the Abelian gluon approximation, 
that we consider are:

a) The pion decay constant $f_\pi$, which is given at first order by the expression \cite{pagelsa,pagelsb}

\be
{\bar{f}}_\pi^2 = \frac{3}{4\pi^2} \int_0^\infty \, dx \, \frac{xM(x)}{[x+M^2(x)]^2}\left( M(x)-\frac{x}{2}\frac{dM(x)}{dx}\right) \,\, .
\label{eq27}
\ee 
There are many improvements to this formula, one of them determined in Ref.\cite{barducci}, which gives the following correction factor
\be
\delta f_\pi^2 =\frac{3}{4\pi^2}\int_0^\infty \, dx \,x^2  \left[x\frac{\left(\frac{dM(x)}{dx}\right)^2-M^2(x)\left(\frac{dM(x)}{dx}\right)^2-M(x)\left(\frac{dM(x)}{dx}\right)}{2[x+M^2(x)]^2}\right]\,\, ,
\label{eq27a}
\ee 
and the pion decay constant $f_\pi$ is obtained as the result of the sum
\be
f_\pi^2 ={\bar{f}}_\pi^2 + \delta f_\pi^2 \,\, ,
\label{eq27b}
\ee
which is to be compared to the experimental value $f_\pi = 93$ MeV \cite{pdg}.

b) The quark condensate $\left\langle {\bar{q}}q\right\rangle$ at the scale $\mu^2 = 1$ GeV$^2$ \cite{robwil}

\be
\left\langle {\bar{q}}q\right\rangle (\mu^2) = - \frac{3}{4\pi^2} \int_0^{\mu^2} \, dx \, \frac{x M(x)}{[x+M^2 (x)]} \,\, ,
\label{eq28}
\ee
which should be compared to typical value of the quark condensate 
$\left\langle {\bar{q}}q\right\rangle$($1$ GeV$^2$) $= \, (229\pm 9)$ MeV$^3$ \cite{gimenez}.

c) The MIT bag constant $\textsl{B}$ \cite{cr}

\be
\textsl{B} = 12 \pi^2 \int_0^\infty \frac{xdx}{(2\pi)^4}\left[ ln\left(\frac{x+M^2(x)}{x}\right)- \frac{M^2(x)}{x+M^2(x)}\right] \,\, ,
\label{eq29}
\ee
which should be compared to the MIT bag constant value of $146$ MeV$^4$ \cite{mita,mitb,mitc}.

The results for the three parameters discussed above, computed as a function of $m$, are shown in Table (\ref{tab1}). Considering
the simple rainbow approximation for the gap equation we see that the results are below the expected values, but better choices for the
vertex function as well as higher order corrections for these quantities will bring them closer to the experimental values. 

\begin{table}
\centerline{%
  \begin{tabular}{|l|c|c|c|c|c|}
    \hline \hline
    \quad $m$ \quad & \quad $\bar{f_{\pi}}$ \quad & \quad $f_{\pi}$\quad  &\quad  $\left\langle \bar{q}q\right\rangle(1\mbox{GeV}^2)$ \quad  & \quad  $B$ \quad \\ 
   \quad  ${[MeV]}$\quad  & \quad ${[MeV]}$\quad & \quad  ${[MeV]}$\quad  & \quad  ${[MeV^3]}$ \quad & \quad ${[MeV^4]}$ \quad \\ \hline 
    160 & 62.0 & 71.12 & 169.03 & 100.09\\ \hline
    180 & 54.51 & 61.83 & 156.12 & 84.96 \\ \hline
    200 & 46.59 & 52.14 & 142.0 & 69.43 \\ \hline
    Expected Values& 93& 93 & $229\pm 9$ &  146  \\ \hline \hline
  \end{tabular}}
\caption{Values of $f_\pi$, $\left\langle {\bar{q}}q\right\rangle$ and $\textsl{B}$ obtained from Eqs.(\ref{eq27})-(\ref{eq29}) as
a function of $m$.}
\label{tab1}  
\end{table}

\section{CSB for higher dimensional representations}

The study of CSB for fermions in higher dimensional representations is of interest because it is a possible way
to verify how this mechanism is distinct from the confinement one, as well as it is important for technicolor model building. 
If a type of Casimir scaling as the one predicted
by Eq.(\ref{cas1}) occurs, we expect that for higher dimensional representations the CSB typical mass scale would
be different from the one for the fundamental representation, and perhaps different from the confinement scale.
It has also been argued that for ``quarks" in the adjoint representation the dynamically massive gluons may have enough
strength to generate a dynamical quark mass \cite{cornwall2,aguilar2}. Indeed in Ref.\cite{aguilar2} a large dynamical
mass was found for fermions in the adjoint representation, and we naively would expect that in this case the confining
and chiral breaking transitions would appear separately. 

If we follow straightforwardly the model
of Ref.\cite{cornwall3} we must also verify what is the difference introduced by the confining propagator in
the case of higher dimensional representations, because in principle we should replace the string tension $K_F$
by $K_R$, which is the string tension for fermions in the representation $R$. We assume that this replacement is
accurate, although we know that the phenomenological potential of Eq.(\ref{pot1}), and consequently the string
tension, does change according to the representation. For instance, in the case of the adjoint representation
it is known that
\be
V_A (r \rightarrow \infty ) = 2 M_g  \,\, ,
\label{pot2}
\ee
where $M_g$ is the energy of the lightest glueball. Moreover, the adjoint representation is not confined but
screened \cite{greensite}, what means that the confining propagator should be understood as effective up to
a certain distance in these cases. Of course, no matter what the fermionic representation is we shall have a
critical distance above which the string breaks. We assume that in the model of Ref.\cite{cornwall3} most of the chiral 
symmetry breaking is still related to the form of Eq.(\ref{eq01}), which does not get the chance to be probed at
large distances, consequently we may still expect that most of
the CSB is driven by the ``confining" propagator. 

The fermion condensate is the most frequent
quantity used to characterize the chiral phase transition, and it is this quantity that we will analyze to
investigate CSB for fermions in different representations.
The fermion condensate is described by Eq.(\ref{eq28}). It is easy to compute this quantity for different fermion
representations if we consider the gap equation in the four-fermion approximation given by Eq.(\ref{eq25}),
perform the angle approximation and neglect the gauge boson mass in the propagator of the $1$-gauge boson exchange contribution:
\br
M_{4f}(p^2)&=&\frac{2}{\pi^3} \frac{K_R}{m^4} \int_0^{m^2} \, d^4k \,  \frac{M_{4f}(k^2)}{k^2+M_{4f}^2(k^2)}  \nonumber \\
&+&\frac{C_2 }{(2\pi)^4}\int \, d^4k \,  \frac{{\bar{g}}^2(p)3M_{4f}(k^2)}{p^2[k^2+M_{4f}^2(k^2)]}\theta (p^2-k^2) \nonumber \\
&+&\frac{C_2 }{(2\pi)^4}\int \, d^4k \,  \frac{{\bar{g}}^2(k)3M_{4f}(k^2)}{k^2[k^2+M_{4f}^2(k^2)]} \theta (k^2-p^2) \,\, .
\label{eq54}
\er
On the other hand we may write the fermion condensate for the representation $R$ in the following form
\be
\left\langle {\bar{q}}q\right\rangle_R (m^2) = - \frac{N_R}{4\pi^2} \int_0^{m^2} \, dx \, \frac{x M_R(x)}{[x+M_R^2 (x)]} \,\, ,
\label{eq55}
\ee
where $N_R$ is the dimension of the fermion representation $R$ and $M_R(x)$ its dynamical mass. 
We are forcing the upper limit of Eq.(\ref{eq55}) to be of ${\cal{O}}(m)$, which
can be as low as $0.18$ GeV, while $\left\langle {\bar{q}}q\right\rangle$ is well known at the $1$ GeV scale.
We did this, as we shall see below, in order to easily compare the condensate expression to Eq.(\ref{eq54}), but
we can check that Eq.(\ref{eq55}) provides a good estimate of the quark condensate at $1$ GeV. To do so we see
that in the four-fermion approximation the dynamical quark mass is almost flat up to $1$ GeV, as can be seen in
Fig.(\ref{fig5}), and its value is around $0.37$, therefore the integral in Eq.(\ref{eq55}) can be approximated
by $\int_0^{1 \,\,GeV^2} dx [0.37x/(x+0.137)] \approx 0.263$. With $N_{R=3}=3$ for the fundamental representation
we obtain $\left\langle {\bar{q}}q\right\rangle = - \left\langle (0.27)^3\right\rangle$ GeV$^3$, which is 
a little large but consistent with the fact that the four-fermion approximation overestimates the dynamical mass.
We stress that the upper limit in the first integral in the right-hand side of Eq.(\ref{eq54}) may be a physical
one in order to be consistent with the critical distance at which the string breaks, as shown in Eq.(\ref{critm}).

Eq.(\ref{eq55}) can be compared to Eq.(\ref{eq54}), the dynamical fermion mass $M_R$ in the four-fermion approximation, 
if we set all integrals at the scale $m$ obtaining
\be
M_R(m^2) \approx \left[ \frac{2K_R}{\pi m^4} + \frac{3C_{2R}g^2_R(m^2)}{16\pi^2 m^2}\right] \int_0^{m^2} \, dx \, \frac{xM_R(x)}{x+M^2_R(x)} \,\, .
\label{eq56}
\ee
Since the first term between brackets in the right-hand side of Eq.(\ref{eq56}) is much larger than the second one, combining the
two last equations we have
\be
\left\langle {\bar{q}}q\right\rangle_R (m^2) \approx - \frac{N_R}{8\pi} \frac{m^4}{K_R} M_R(m^2) \,\, .
\label{eq57}
\ee 
In the QCD case, for quarks in the fundamental
representation, this relation underestimates the condensate due to the fact that the integration area in this
equation was drastically reduced when we cutoff the integral at $m^2$, whereas the dynamical quark mass solution,
as shown in Fig.(\ref{fig5}), is almost flat up to $1$ GeV$^2$. Since the effect of the Eq.(\ref{eq01}) is the dominating one, it
is quite plausible that the relation of Eq.(\ref{eq57}) holds up to other scales (still keeping the factor $m^4$ in
the right-hand side) and it can be tested through lattice simulations. 

We can now make a few comments on the differences between CSB for fermions in the fundamental and adjoint representations estimating the ratio of the condensates for $SU(3)$ fermions in these representations. First
we need to know how the string tension changes as we change the fermion representation.
It has been observed in lattice simulations what is usually called Casimir scaling for the string tension \cite{greensite}, i.e.
\be
K_R \approx \frac{C_R}{C_F} K_F   \,\, ,
\label{eq51}
\ee 
where $C_R/C_F$ is the ratio between the Casimir operators for the representation $R$ and the fundamental one.
For $SU(N)$ theories and a finite $N$ the Casimir scaling law must break down at some point, to be replaced by a dependence
on the $N$-ality $k$ of the representation \cite{greensite}
\be
K_R = f(k) K_F 
\label{eq52}
\ee
This change of behavior is credited to an effect of force screening by the gauge bosons. For fermions in the adjoint representation
the $N$-ality is zero, therefore, according to Casimir scaling, the adjoint string tension is given by
\be
K_A = \frac{2N^2}{N^2-1} K_F \,\, ,
\label{eq53}
\ee
and, as a reasonable approximation, we may assume $K_A \approx 2K_F$. 
Consequently we obtain the following ratio at the scale $m^2$
\be
\frac{\left\langle {\bar{q}}q\right\rangle_3 }{\left\langle {\bar{q}}q\right\rangle_8 } \approx \frac{3}{4} \frac{M_3}{M_8} \,\, .
\label{eq58}
\ee
Once the dynamical masses almost scale with the string tension value we could say that the above ratio is roughly of order $3/8$.
Of course, the uncertainty in this estimative is certainly connected to the remarks made at the beginning of this section about
the phenomenological potential and the effective propagator for the adjoint representation. For other fermionic representations
the screening behavior is smaller, although in all cases we certainly have a limit on the critical distance for which this approach
is valid that will be connected with the string breaking mechanism.  

\section{Conclusions}

In this work chiral symmetry breaking for QCD-like theories was studied in the case of a rainbow Schwinger-Dyson equation in the
presence of dynamically massive gauge bosons. Confinement was also introduced into the SDE in the form of an effective propagator
which is one that can reproduce an area law for quarks. 
This is a phenomenological way to investigate the possibility, indicated by the lattice, of confinement (by center vortices)
being intrinsically related to chiral symmetry breaking.

We briefly review the conditions for the confining propagator to generate non-trivial massive solutions. We studied the bifurcation
condition for the complete gap equation, i.e. the one with the exchange of dynamically massive gauge bosons and with the
inclusion of the confining effective propagator dependent on the entropic parameter $m$, which must be proportional
to the dynamical quark mass, and on the string tension $K_F$, verifying that there is a maximum $m$ value
below which the chiral symmetry is broken, and generating the expected values for the dynamical quark masses. As already known in
the literature we verified that the massive gauge boson exchange gives only a minute contribution to the dynamical fermion mass.
Most of the breaking is due to the confining propagator and this may be one indication that we should not expect large differences
between the chiral and confinement transitions. 

The asymptotic behavior of the gap equation was studied in order to evaluate how confinement may affect the asymptotic self-energy.
We observed that if the confining effect is restricted to a small momentum region, presenting some arguments for why
this may happen, the solution changes its behavior from the one
obtained with the help of the operator product expansion to a self-energy typical of a four-fermion interaction. It is known
that the massive gauge boson gap equation cannot be reduced to an effective four-fermion interaction due to the small infrared strength
of the product ``coupling$\otimes$propagator", however we verified that this is not the case for the confining effective propagator. The
bifurcation equation for the gap equation with the four-fermion approximation performed in the confining part reproduces
approximately the result of the complete gap equation without any approximation. Actually this approximation overestimates the dynamically
generated fermion mass by an  ${\cal{O}}(10\%)$.

The complete gap equation was computed numerically without any approximation in Section V. Our result for the dynamical
fermion mass is consistent with the one obtained by Cornwall \cite{cornwall3}. We also computed several chiral parameters
which resulted to be of the order of the expected experimental values. These parameters show dependence on the effective propagator
scale $m$, however larger differences in the dynamical mass can also be observed with a small variation of the string tension
value (see Fig.(\ref{fig3l})).

In Section VI we discuss the case of chiral symmetry breaking for fermionic representations different than the fundamental one.
We found a simple relation between the fermion condensate and the fermionic dynamical mass. This relation depends on the
dimension of the fermion representation, $m$ and $K_F$, and we assumed that the form of Eq.(\ref{eq01}) still holds for
different fermionic representations up to a certain distance. In principle such relation can be studied in lattice simulations. Since most
of the symmetry breaking is due to the confining effective propagator and the string tension approximately scales with the Casimir
operator, we do expect simple relations between the condensate values for different representations, but no matter what
representation we choose it seems that in this scenario we shall not have large differences between the confining and chiral transition mass scales.

Our results indicate that the CSB mechanism proposed in Ref.\cite{cornwall3} can account for the expected values of several
known chiral parameters. The model also seems to indicate that the CSB scale is connected to the confinement one even
for fermions in higher dimensional representations.
How far can we assume the four-fermion approximation discussed here for the purpose of practical 
calculations still needs further analysis. Finally, a more precise determination of the CSB observables can be obtained with 
the introduction of more sophisticated vertex functions and higher order corrections. 
\\
\\
\noindent{\bf Acknowledgments}
\\ 
\par We are indebted to A. C. Aguilar for discussions and help with the numerical calculation and are grateful to Prof. J. M. Cornwall for 
a clarifying exchange of correspondence and comments on a previous version of this manuscript. This research was partially supported by 
the Conselho Nacional de Desenvolvimento Cient\'{\i}fico e Tecnol\'ogico (CNPq) (AD and AAN), and Funda\c c\~ao de Amparo a Pesquisa do Estado de S\~ao Paulo 
(FAPESP) (FAM).

\end{document}